# A Biophysical Approach to Production Theory

Jing Chen
School of Business
University of Northern British Columbia
Prince George, BC
Canada V2N 4Z9
Phone: 1-250-960-6480
Email: chenj@unbc.ca
Web: http://web.unbc.ca/~chenj/

James K. Galbraith
Lyndon B. Johnson School of Public Affairs
The University of Texas at Austin
Austin, TX 78713-8925, USA
Phone: 512-471-1244
Email: galbraith@mail.utexas.edu
Web: http://www.utexas.edu/lbj/faculty/james-galbraith/

Preliminary draft. Comments welcome. We thank Sufey Chen and seminar participants at University of Texas, Austin for helpful comments.

## Abstract

Most people agree that human activities are consistent with physical laws. One may naturally think that sensible economic theories can be derived from physical laws and evolutionary principles. This is indeed the case. In this paper, we present a newly developed production theory of economics from biophysical principles. The theory is a compact analytical model that provides, in our view, a much more realistic understanding of economic (as well as social and biological) phenomena than the neoclassical theory of production.

## 1. Introduction

Most people agree that human activities are consistent with physical laws. One may naturally think that sensible economic theories can be derived from physical laws and evolutionary principles. This is indeed the case. A newly developed analytical thermodynamic theory of economics provides a much more realistic and intuitive understanding of economic, social and biological phenomena than the mainstream economic theory (Chen, 2005). In this paper, we will present a more detailed discussion of the production theory, which forms one part of the analytical theory. We will show how the new production theory provides a simple and systematic understanding about investment activities and economic policies.

There are two fundamental properties of life. First, living organisms need to extract resources from the environment to compensate for the continuous diffusion of resources required to maintain various functions of life. Second, for an organism to be viable, the total cost of extracting resources has to be less than the amount of resources extracted (Odum, 1971 and Hall et al, 1986; Rees, 1992). The second property could be understood as natural selection rephrased from the resource perspective. The purpose of our work is to show that a self-contained mathematical theory of production can be derived from these two fundamental properties of life.

From the physics perspective, resources can be regarded as low entropy sources (Georgescu-Roegen, 1971). The entropy law states that systems tend toward higher entropy states spontaneously. Living systems, as non-equilibrium systems, need to extract low entropy source from the environment to compensate for their continuous dissipation. It can be represented mathematically by lognormal processes, which contain both a growth term and a dissipation term. According to the entropy law, the thermodynamic dissipation of an organic or economic system is spontaneous. However the extraction of low entropy source from the environment depends on specific biological or institutional structures that incur fixed or maintenance costs. Additional variable cost is also required for resource extraction. Higher fixed cost systems generally have lower variable costs. Fixed cost is largely determined by the genetic structure of an organism or the design of a project. Variable cost is a function of the environment. From the Feynman-Kac formula, a result widely used in science literature and increasingly used in finance literature, we derive the equation that variable cost of a production system should satisfy. An organism survives if the amount of resources it extracts is higher than the total cost spent. Similarly, a business survives if its revenue is higher than the total cost of production. We set the initial condition of the equation so that total cost is equal to the amount of resource extracted or revenue generated. Then we solve the equation to derive a formula of variable cost as a mathematical function of product value, fixed cost, uncertainty, discount rate and project duration. From this formula of variable cost, together with fixed cost and volume of output, we can compute and analyze the returns and profits of different production systems under various kinds of environment in a simple and systematic way. The results are highly consistent with the empirical evidences obtained from the vast amount of literature in economics and ecology. Furthermore, by putting major factors of production into a compact mathematical model, the theory provides precise insights about the tradeoffs and constraints of various business or evolutionary strategies that are often lost in intuitive thinking.

Product value, fixed cost, variable cost, discount rate, uncertainty, project duration and volume of output are major factors in production. These factors naturally became the center of investigation in early economic literature. However, because of the difficulty in forming a compact mathematical model about these factors, discussion about these factors becomes peripheral in current economic literature. With the help of this analytical production theory, theoretical investigation in economics may refocus on important issues in economic activities.

Among the various relations of different factors, the relation between fixed cost and variable cost is probably the most important. We will elaborate on this relation further. People observed that useful energy comes from the differential or gradient between two parts of a system. In general, the higher the differential between two parts of a system, the more efficient the work becomes. At the same time, it is more difficult to maintain a system with high differential. In other words, a lower variable cost system requires higher fixed cost to maintain it. This is a general principle. We can list several familiar examples from physics and engineering, biology and economics.

In an internal combustion engine, the higher the temperature differential between the combustion chamber and the environment, the higher the efficiency in transforming heat into work. This is the famed Carnot's Principle, the foundation of thermodynamics. At the same time, it is more expensive to build a combustion chamber that can withstand higher temperature and pressure. Diesel burns at higher temperature than gasoline. This is why the energy efficiency of diesel engine is higher and the cost of building a diesel engine is higher. In electricity transmission, higher voltage will lower heat loss. At the same time, higher voltage transmission systems are more expensive to build and maintain because the distance from the line to the ground has to be longer to reduce the risk of electric shock. The differential of water levels inside and outside a hydro dam generates electricity. The higher the hydro dam, the more electricity can be generated. At the same time, a higher hydro dam is more costly to build and maintain. Warm blooded animal can run faster than cold blooded animals because their body temperature is maintained at high levels to ensure fast biochemical reactions. But the basic metabolism rates of warm blooded animals are much higher than the cold blooded animals. Shops located near high traffic flows generate high sales volume per unit time. But the rent costs in such locations are also higher. Well trained employees work more efficiently. But employee training is costly. The tradeoff between lower variable cost and higher fixed cost is often not explicitly discussed in the same literature.

However, the new production theory provides a quantitative measure of return and profit at different levels of fixed cost and variable cost under different kinds of circumstances.

From the production theory, it can be calculated that when the fixed cost is zero, the variable cost is equal to the product value. This means that any organisms or organizations need to make fixed investment before earning a positive return. This simple result has broad implications. It means that any viable organization, whether a company or a country, needs a common fixed investment. We will discuss tax policy and the concept of market from the perspective of fixed cost and variable cost further.

In current economic literature, taxation is often described as a type of distortion or imperfection. If this is true, any society that abolishes taxation will remove the distortion and becomes less imperfect. It will crowd out other social systems that collect taxes. However, this has not happened. Montesquieu (1748) observed long ago, "In moderate states, there is a compensation for heavy taxes; it is liberty. In despotic states, there is an equivalent for liberty; it is the modest taxes." From the mainstream economic theory, we might conclude that despotic states are more perfect than moderate states.

In this production theory, tax is considered as a fixed cost of the whole society. Therefore, lower tax rate does not automatically mean a better society. The proper level of taxation should be determined by other factors in the society, such as the general living standard of the population.

In neoclassical economics, market is a very abstract concept.

> The "market" in modern usage is not some physical location. … The market is the nonstate, and thus it can do everything the state can do but with none of the procedures or

> rules or limitations. … Because the word lacks any observable, regular, consistent meaning, marvelous powers can be assigned. The market establishes Value. It resolves conflict. It ensures Efficiency in the assignment of each factor of production to its most Valued use. … From each according to Supply, to each according to Demand. The market is thus truly a type of God, "wiser and more powerful than the largest computer," … Markets achieve effortlessly exactly what governments fail to achieve by directed effort. (Galbraith, 2008, p. 20)

In this production theory, market is a concrete concept. The structure of market is determined by its various parameters. If a market is very small, then its structure will be very simple to reduce fixed cost. If a market is very large, then its structure will be very sophisticated to reduce variable cost. For example, a village market has little formal structure or regulation. But New York stock exchange has very expensive computer systems and highly complex regulatory structures to ensure the smooth flow of large volume of transactions. In this concrete structure, it is meaningless to discuss whether market is "efficient" or not. Any market that turns a positive return will survive and prosper. Any market that turns a negative return will shrink and disappear.

For an organism or organization, part of its fixed cost is used to regulate its internal environment to maintain a proper working condition. For example, the human body is regulated at around 37 degree centigrade. When a person is infected, the body temperature is regulated to higher temperature to more effectively attack the infecting bacteria. However, temperature that is too high will damage the brain's information processing capability, which is sensitive to thermal noises (Gisolfi, Carl and Mora, Francisco, 2000). Hence, regulation is a compromise between different parts of an organism. When a part of an organism escapes regulation and turns into a free market, that part of the organism will grow rapidly. This is called cancer. In the end, the unregulated growth, if it is not stopped, will drain all the resources of the organism and destroy the organism. In human society, if an economic sector gains control of large amount of resources and escapes regulation, it will grow rapidly and generate huge profit for the insiders. But the process will drain resources from the whole society. This is what we have witnessed in the current financial crisis.

Mainstream economists pay little attention to the biophysical foundation of human society because they believe that the special qualities of human beings will make physical constraints less relevant. However, thinking about constraints is a very effective way to understand social phenomena (Galbraith, 2008). A biophysical approach puts the physical constraint of human society at the center of its analysis. The validity of a physical theory of economics is best manifested by the existence of a corresponding mathematical theory that is derived from the most fundamental properties of life and is consistent with a wide range of patterns observed in both economics and ecology. After all, all physical laws are represented by mathematical formulas. By establishing an analytical theory based on biophysical principles, many philosophical and verbal problems can be turned into scientific and quantitative inquiries. The computability of the mathematical theory will transform biological science, which include social science as a special case, into an integral part of physics.

This paper is an update from earlier works (Chen, 2005, 2008). It is structured as follows. Section two presents the derivation of the production theory. Section three presents detailed results from this theory. Section four concludes.

## 2. An analytical thermodynamic theory of production

The theory described in this section can be applied to both biological and economic systems. For simplicity of exposition, we will use the language of economics. However the extension to biological system is straight forward.

A basic property in economic activities is uncertainty. While a business may face many different kinds of uncertainty, most of the uncertainties are reflected in the price uncertainty of the product. Suppose $S$ represents unit price of a commodity, $r$, the expected rate of change of price and σ, the rate of uncertainty. Then the process of $S$ can be represented by the lognormal process

$$\frac{dS}{S} = rdt + \sigma dz \,. \tag{1}$$

where

$dz = \varepsilon\sqrt{dt}, \quad \varepsilon \in N(0,1)$ is a random varaible with standard Gaussian distribution

The production of the commodity involves fixed cost and variable cost. In general, production factors that last for a long term, such as capital equipment, are considered fixed cost while production factors that last for a short term, such as raw materials, are considered variable costs. If employees are on long term contracts, they may be better classified as fixed costs, although in many cases, they are classified as variable costs. Firms can adjust their level of fixed and variable costs to achieve high level of return on their investment. Intuitively, in a large and stable market, firms will invest heavily on fixed cost to reduce variable cost, thus achieving a higher level of economy of scale. In a small or volatile market, firms will invest less on fixed cost to maintain high level of flexibility. In the following, we will derive a formal mathematical theory that focuses on this issue.

In natural science, there is a long tradition of studying stochastic processes with deterministic partial differential equations. For example, heat is a random movement of molecules. Yet the heat process is often studied by using heat equations, a type of partial differential equations. In studying quantum electrodynamics, Richard Feynman (1948) developed a general method to study probability wave function with partial differential equations. Kac (1951) provided a more systematic exposition of this method, which was later known as the Feynman-Kac formula, whose use is very common in natural sciences (Kac, 1985). Recently, the Feynman-Kac formula has been widely used in the research in finance. Our goal is to apply the Feynman-Kac formula to derive variable cost in production as a function of other parameters.

Let *K* represent fixed cost and *C* represent variable cost, which is a function of *S*, the value of the commodity. According to the Feynman-Kac formula (Øksendal, 1998, p. 135), if the discount rate of a firm is *r*, the variable cost, *C*, as a function of *S*, satisfies the following equation

$$\frac{\partial C}{\partial t} = rS\frac{\partial C}{\partial S} + \frac{1}{2}\sigma^2 S^2 \frac{\partial^2 C}{\partial S^2} - rC \tag{2}$$

with the initial condition

$$C(S,0) = f(S) \tag{3}$$

To determine *f(S),* we perform a thought experiment about a project with a duration that is infinitesimally small. When the duration of a project is sufficiently small, it only has enough time to produce one unit of product. In this situation, if the fixed cost is lower than the value of the product, the variable cost should be the difference between the value of the product and the fixed cost to avoid arbitrage opportunity. If the fixed cost is higher than the value of the product, there should be no extra variable cost needed for this product. Mathematically, the initial condition for the variable cost is the following:

$$C(S,0) = \max(S - K, 0) \tag{4}$$

where $S$ is the value of the commodity and $K$ is the fixed cost of a project. When the duration of a project is $T$, solving equation (2) with the initial condition (4) yields the following solution

$$C = SN(d_1) - Ke^{-rT}N(d_2) \tag{5}$$

where

$$d_1 = \frac{\ln(S/K) + (r + \sigma^2/2)T}{\sigma\sqrt{T}}$$
$$d_2 = \frac{\ln(S/K) + (r - \sigma^2/2)T}{\sigma\sqrt{T}} = d_1 - \sigma\sqrt{T}$$

The function *N(x)* is the cumulative probability distribution function for a standardized normal random variable. Formula (5) takes the same form as the well-known Black-Scholes (1973) formula for European call options.

Formula (5) provides an analytical formula of variable cost as a function of product value, fixed cost, uncertainty, duration of project and discount rate of a firm. Similar to understanding in physics, the calculated variable cost is the average expected cost (Kiselev, Shnir, Tregubovich, 2000; Tong, 2008).

We will briefly examine the properties of this formula. First, the variable cost is always less the value of the product when the fixed cost is positive. No one will invest in a project if the expected variable cost is higher than the product value. Second, when the fixed cost is zero, the variable cost is equal to the value of the product. When the fixed cost approaches zero, the variable cost will approach the value of the product. This means that businesses need fixed investment before they can make a profit. Similarly, any organisms need to invest in a fixed structure before extracting resources profitably. Third, when fixed costs, $K$, are higher, variable costs, $C$, are lower. Fourth, for the same amount of the fixed cost, when the duration of a project, $T$, is longer, the variable cost is higher. Fifth, when uncertainty, $\sigma$, increases, the variable cost increases. Sixth, when discount rate becomes lower, the variable cost decreases. This is due to the lower cost of borrowing. All these properties are consistent with our intuitive understanding of production processes.

Suppose the volume of output during the project life is $Q$, which is bound by production capacity or market size. We assume the present value of the product to be $S$ and the variable cost to be $C$ during the project life. Then the total present value of the product and the total cost of production are

$$SQ \text{ and } CQ + K \tag{6}$$

respectively. The return of this project can be represented by

$$\ln\left(\frac{SQ}{CQ+K}\right) \tag{7}$$

and the net present value of the project is

$$QS-(QC+K)=Q(S-C)-K \tag{8}$$

Unlike a conceptual framework, this mathematical theory enables us to make quantitative calculations of returns of different projects under different kinds of environments. In the next section we will provide a systematic analysis.

## 3. Systematic analysis of the performance of a project

The profit or return of a project is determined by fixed cost, variable cost and total output during the life of the project. Variable cost is a function of product value, fixed cost, uncertainty, duration of the project, and discount rate. At the project level, how much to invest or commit to at the beginning of the project is the most important decision to make. At the country level, how much to invest in roads, public schools, libraries and other public assets is often the defining characteristic of a state. At the project level, the cost of borrowing often determines the viability or structure of a project. At the country level, the main tool for many central banks to fine tune economic activities is by adjusting discount rates. Biological species are often classified as *K* and *r* species. In general, *K* species have high fixed investment and *r* species have high discount rate. Therefore, fixed cost and discount rate are often the most important factors in investment decision making. In this section, we will discuss how different factors are related to each other in affecting the performance of investment projects. We will group the factors first by fixed cost and then by discount rate.

***Fixed cost and discount rate:***

We discuss how the level of fixed cost affects the preference for discount rates. Assume there are two production systems, one with a fixed cost of 10 and the other with a fixed cost of 5. Other parameters of the production systems are the same. The unit value of the product is 1, the duration of the projects is 10 years and the level of uncertainty is 60% per annum. We will calculate how variable costs change with different discount rates. When discount rates are decreased, the variable costs of high fixed cost systems decrease faster than the variable costs of low fixed cost systems (Figure 1). This indicates that high fixed cost systems have more incentive to maintain low discount rates or lending rates. It helps us understand why prevailing lending rates are different at different areas or times.

In medieval societies or in less developed countries, lending rates were very high; in modern economy, lending rates charged by regular financial institutions are generally very low. To maintain a low level of lending rates, many credit and legal agencies are needed to inform and enforce, which is very costly. As modern societies are of high fixed cost, they are willing to put up the high cost of credit and legal agencies because the efficiency gain from lower lending rate is higher in high fixed cost systems.

While this result about fixed cost, discount rate and variable cost seems to be new, the human mind understands it instinctly.  In the field of human psychology, there is an empirical regularity called the “magnitude effect” (small outcomes are discounted more than large ones) Most studies that vary outcome size have found that large outcomes are discounted at a lower rate than small ones. In Thaler's (1981) study, respondents were, on average, indifferent between $15 immediately and $60 in a year, $250 immediately and $350 in a year, and $3000 immediately and $4,000 in a year, implying discount rates of 139%, 34% and 29%, respectively (Frederick, Loewenstein and O'Donoghue, 2004). This shows that the human mind intuitively understands the relation between discount rate and variable cost at different level of assets.

Differences in fixed costs in child bearing between women and men also affect the differences in discount rates between them. Women spend much more effort in child bearing. From our theory, the high fixed investment women put in child bearing would make women's discount rate lower than men's. An informal survey conducted in a classroom survey showed that discount rates of the female students are lower than that of the male students.

***Fixed cost and uncertainty:***

By calculating variable costs from (5), we find that, as fixed costs are increased, variable costs decrease rapidly in a low uncertainty environment and change very little in a high uncertainty environment. To put it in another way, high fixed cost systems are very sensitive to the change of uncertainty level while low fixed cost systems are not. This is illustrated in Figure 2.

The above calculation indicates that systems with higher fixed investment are more effective in a low uncertainty environment and systems with lower fixed investment are more flexible in high uncertainty environment. This explains why mature industries, such as household supplies, are dominated by large companies such as P&G while innovative industries, such as IT, are pioneered by small and new firms. Microsoft, Apple, Yahoo, Google, Facebook and countless other innovative businesses are started by one or two individuals and not by established firms.

Similarly, in scientific research, mature areas are generally dominated by top researchers from elite schools, while scientific revolutions are often initiated by newcomers or outsiders (Kuhn, 1996).

***Fixed cost and duration of the project:***

We will study how the level of the duration of projects affects the rate of return. If the duration of a project is too short, we may not be able to recoup the fixed cost invested in the project. If the duration of a project is too long, the variable cost, or the maintenance cost may become too high. This is a natural tradeoff between duration and maintenance cost. With the mathematical theory, we can make quantitative calculations. To be specific, we will compare the profit level of one project with that of two projects with duration half long while keeping other parameters identical. We also assume the annual output of two types of projects are the same. We find that when duration is short, the profit level of one long project is higher than two short projects. When duration is long, the profit level of one project is lower than two short projects. This is consistent with intuition. The detailed calculation is illustrated in Figure 3. It explains why individual life does not go on forever. Instead, it is of higher return for animals to have finite life span and produce offspring. This also explains why most businesses fail in the end (Ormerod, 2005).

Calculation also shows that when the level of fixed cost increases, the length of duration for a project to be of positive return also increases. This suggests that large animals and large projects, which have higher fixed cost, often have longer life. There is an empirical regularity that animals of larger sizes generally live longer (Whitfield, 2006). The relation between fixed cost and duration can be also applied to human relation. In child bearing, women spend much more effort than men. Therefore we would expect women value long term relation while men often seek short term relation, which is indeed the case most of the time (Pinker, 1997).

***Fixed cost and the volume of output or market size:***

We now discuss the returns of investment on projects of different fixed costs with respect to the volume of output or market size. Figure 4 is the graphic representation of (7) for different levels of fixed costs. In general, higher fixed cost projects need higher output volume to breakeven. At the same time, higher fixed cost projects, which have lower variable costs in production, earn higher rates of return in large markets.

We can see from the above discussion that the proper level of fixed investment in a project depends on the expectation of the level of uncertainty and the size of the market. When the outlook is stable and the market size is large, projects with high fixed investment earn higher rates of return. When the outlook is uncertain or market size is small, projects with low fixed cost breakeven easier.

In the ecological system, the market size can be understood as the size of resource base. When resources are abundant, an ecological system can support large, complex organisms (Colinvaux, 1978). Physicists and biologists are often puzzled by the apparent tendency for biological systems to form complex structures, which seems to contradict the second law of thermodynamics (Schneider and Sagan, 2005; Rubí, 2008). However, once we realize that systems of higher fixed cost are more competitive in the resource rich and stable environments, this evolutionary pattern becomes easy to understand.

***Discount rate and project duration:***

When the discount rate or interest rate becomes lower, the variable cost of a project will decrease and profit will increase. Projects with different lengths of duration will be affected differently from the reduction of discount rates. Figure 5 presents the ratios of profits between projects at low and high discount rates at different levels of project duration. As project lengths are increased, the ratios increase as well. This indicates that projects with longer duration benefit more from the reduction of interest rates.

Next we calculate the breakeven point of a project with respect to the project duration and the discount rate. Let us assume that project output per unit of time is one. The calculation from formula (7) shows that it requires lower discount rate to breakeven when the project duration is lengthened. When the project duration is sufficiently long, the discount rate could become negative at the breakeven point. This is illustrated in Figure 6. In ecological systems, animals that live a long life, such as elephants, generally have much lower fertility rates than animals that live a short life, such as cats. In human society, we often use longevity, or duration of human life as an indicator of the quality of a social environment. At the same time, societies that enjoy a long life span, such as Japan, are often concerned about below replacement fertility. Our calculation shows that there is a negative correlation between duration of life and fertility. Intuitively, the aging population needs a great amount of resources to maintain their health, which reduces the amount of resources available to support children. Hence, there is a natural tradeoff between longevity and fertility.

***Discount rate and uncertainty***

Variable cost is an increasing function of discount rate. When uncertainty is low, variable cost is much lower with a low level of discount rate. When uncertainty is high, variable costs are not sensitive to discount rate. Therefore, it is only important to reduce discount rate in a stable environment. Figure 7 presents the ratios of variable costs between low and high levels of uncertainty at different levels of discount rates. It shows that the reduction of the variable cost is much more significant at a low uncertainty level. This explains why *r* species, which have high discount rates, often thrive in highly uncertain environments.

***Volume of output and uncertainty***

In the earlier part of this paper, we assumed that the volume of the output of a company does not affect other factors in production. However, when the size of a company increases and the business expands, the internal coordination and external marketing becomes more complex. This can be modeled with uncertainty, $\sigma$, as an increasing function of the volume of the output. Specifically, we can assume

$$\sigma = \sigma_0 + lQ$$

Where $\sigma_0$ is the base level of uncertainty, $Q$ is the volume of output and $l > 0$ is a coefficient. With the new assumption, we can calculate the return of production from formula (7). The result from the calculation is presented in Figure 8. From Figure 8, the rate of return initially increases with the production scale, which is well known as the economy of scale. When the size of the output increases further, the rate of return begin to decline. This is the law of diminishing return.

***An application: Understanding the effects of monetary policy***

Central banks often adjust the levels of the interest rate to fine tune economic growth rates. With this analytical production theory, we can clearly understand the effects of monetary policies. According to our early analysis, when interest rates are lower, investments with high fixed cost and long duration will benefit more. Since loans are fixed income instruments for buyers, they are fixed costs for issuers (Chen, 2006). Hence, mortgages with long maturity and low initial payment are long duration, high fixed cost investments. They tend to do well in a low interest rate environment. With the burst of stock market bubble in early 2000, interest rates were lowered by the central banks to stimulate economy. While the monetary policy was aimed to help economy in general, low down payment and long duration mortgage businesses will benefit most and hence attract large amounts of capital. This is the source of financial crisis induced by subprime mortgages.

## 4. Concluding remarks

In this paper, we present an analytical theory of production derived from the biophysical principles. The generality of this theory is a consequence from the generality of physical laws. Physical systems, biological systems and economic systems all follow the same natural laws. This allows us to develop a unified production theory that can be applied to many different fields. In particular, the economic theory provides a new way to understand physical phenomena. Some problems in non-equilibrium thermodynamics have puzzled researchers for a long time (Rubí, 2008). They can be easily resolved from the economic principle developed in this theory. Historically, some economic type principles in physics, such as principle of least action and maximum entropy principle (Jaynes, 1957), have been very fruitful in providing unified foundations to very diverse areas of investigation. The production theory presented in this work has provided a unified understanding for a wide range of problems in economics and biology.

**Figure captions**

**Figure 1. Fixed cost and discount rate:** When discount rates are decreased, variable costs of high fixed cost systems decreases faster than variable costs of low fixed cost systems.

**Figure 2. Fixed cost and uncertainty**: In a low uncertainty environment, variable cost drops sharply as fixed costs are increased. In a high uncertainty environment, variable costs change little with the level of fixed cost.

**Figure 3. Fixed cost and duration of the project:** Comparison of the profit level of one project with that of two projects with duration half long while keeping other parameters identical. Assume the annual output of two types of projects are the same. When duration is short, the profit level of one long project is higher than two short projects. When duration is long, the profit level of one project is lower than two short projects.

**Figure 4. Fixed cost and the volume of output**: For a large fixed cost investment, the breakeven market size is higher and the return curve is steeper. The opposite is true for a small fixed cost investment.

**Figure 5. Project duration and discount rate:** the ratios of profits between projects at low and high discount rates at different levels of project duration

**Figure 6. Longevity and population growth rate:** The tradeoff between longevity and population growth

**Figure 7. Uncertainty and discount rate:** the ratios of variable costs between low and high levels of uncertainty at different levels of discount rates

**Figure 8. Volume of output and the rate of return:** The rate of return of a project with respect to volume of output, when diffusion is an increasing function of volume of output

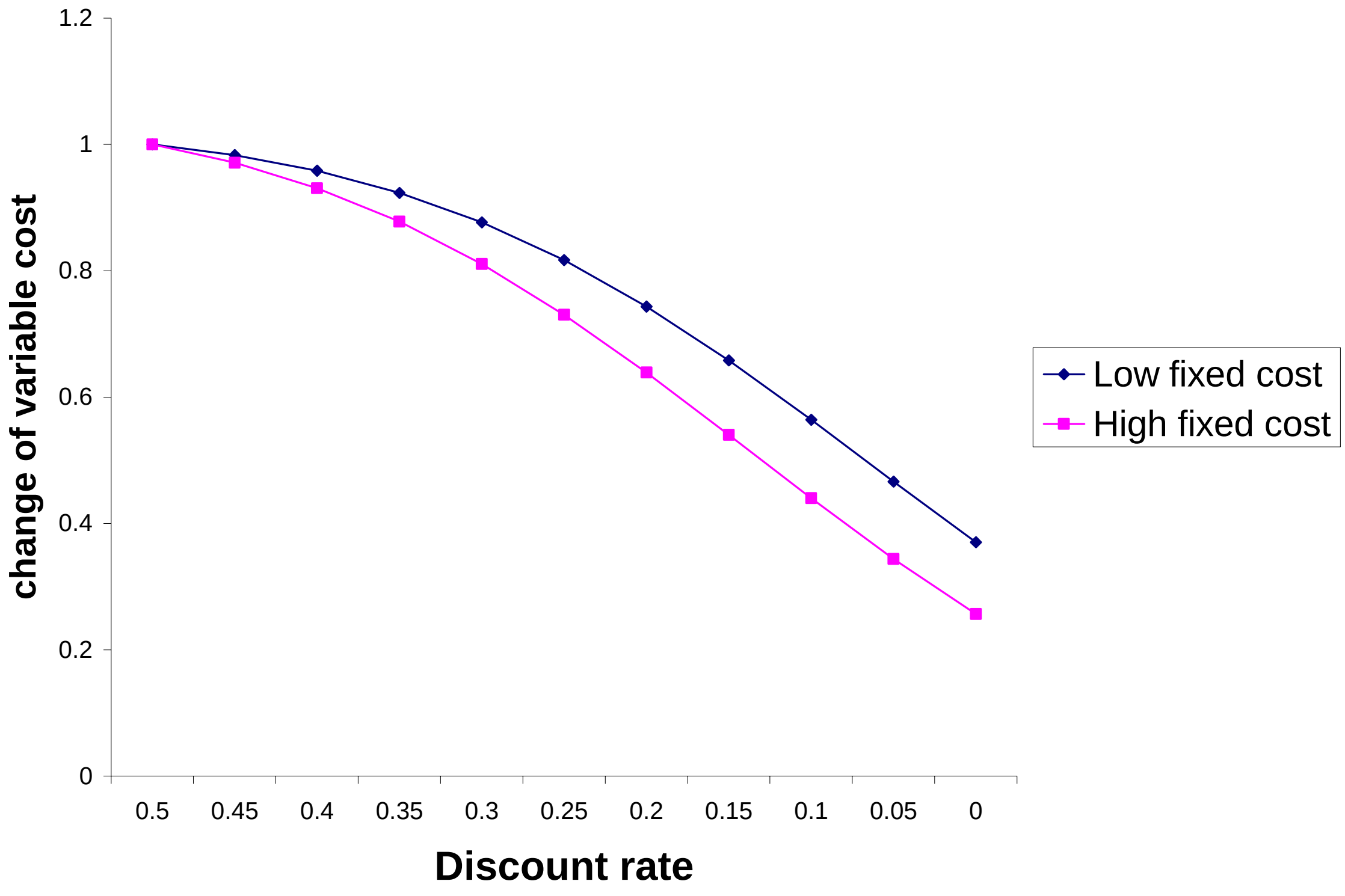

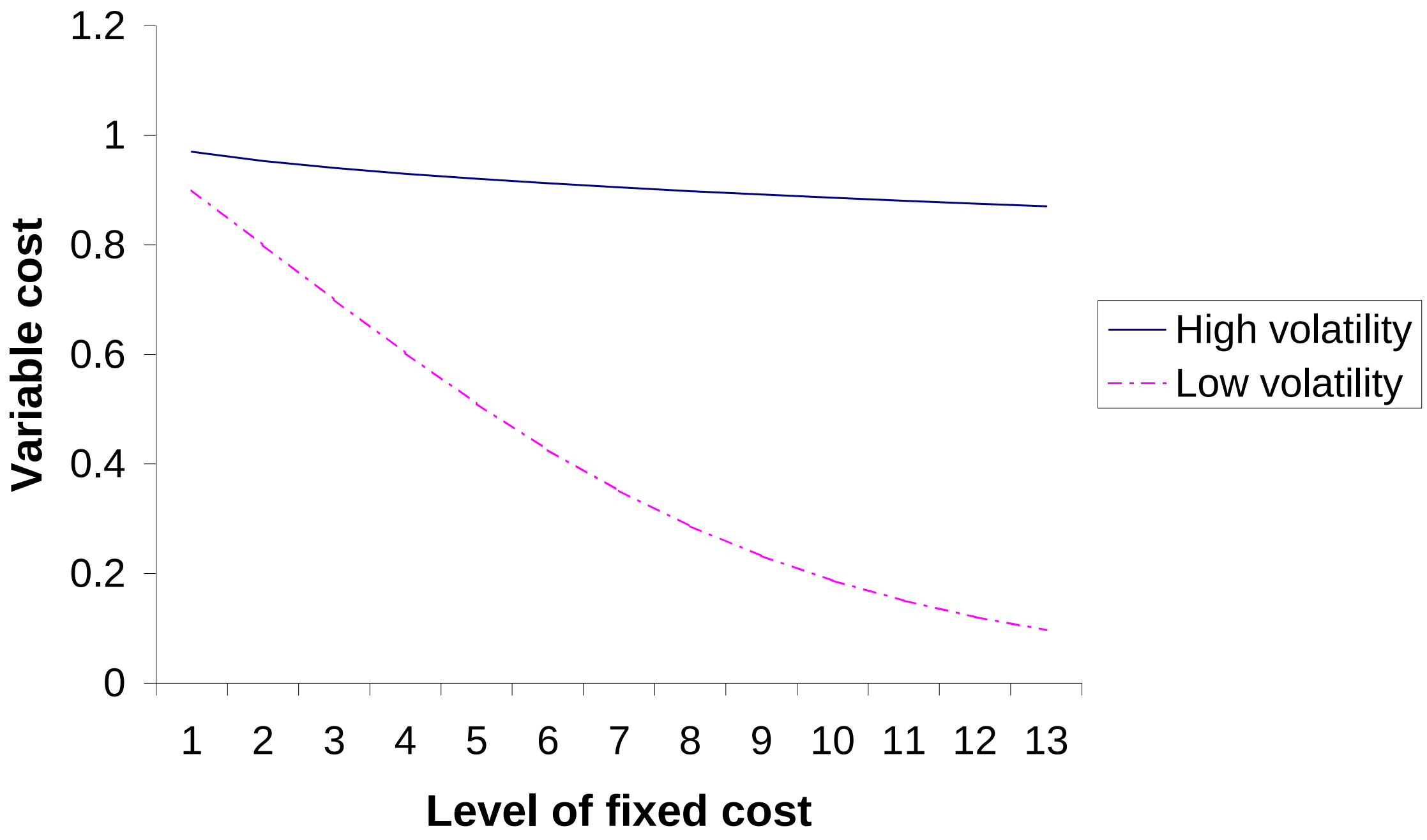
1.2
1
0.8
0.6
0.4
0.2
0
Variable cost
1 2 3 4 5 6 7 8 9 10 11 12 13
Level of fixed cost
High volatility
Low volatility

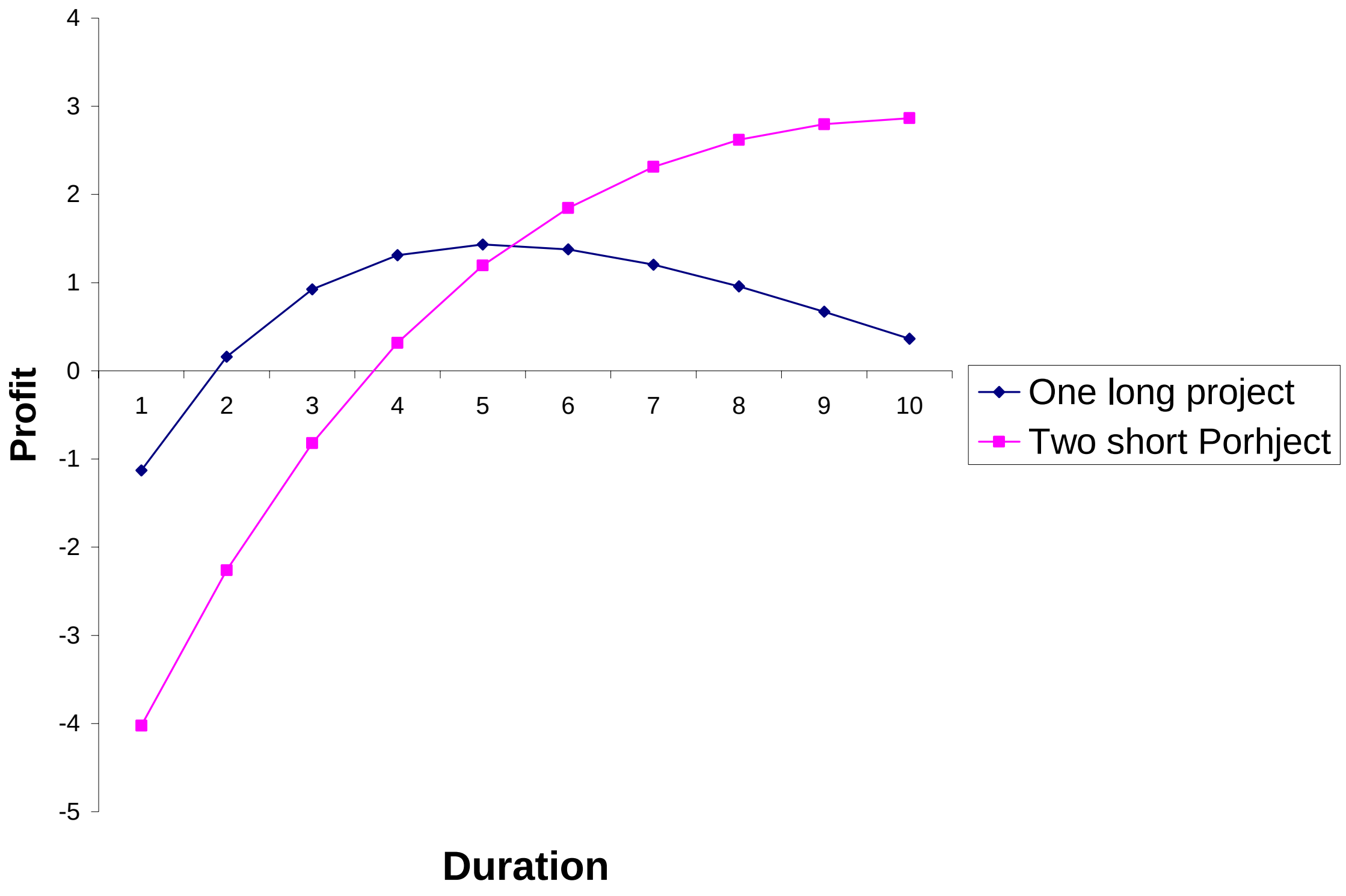

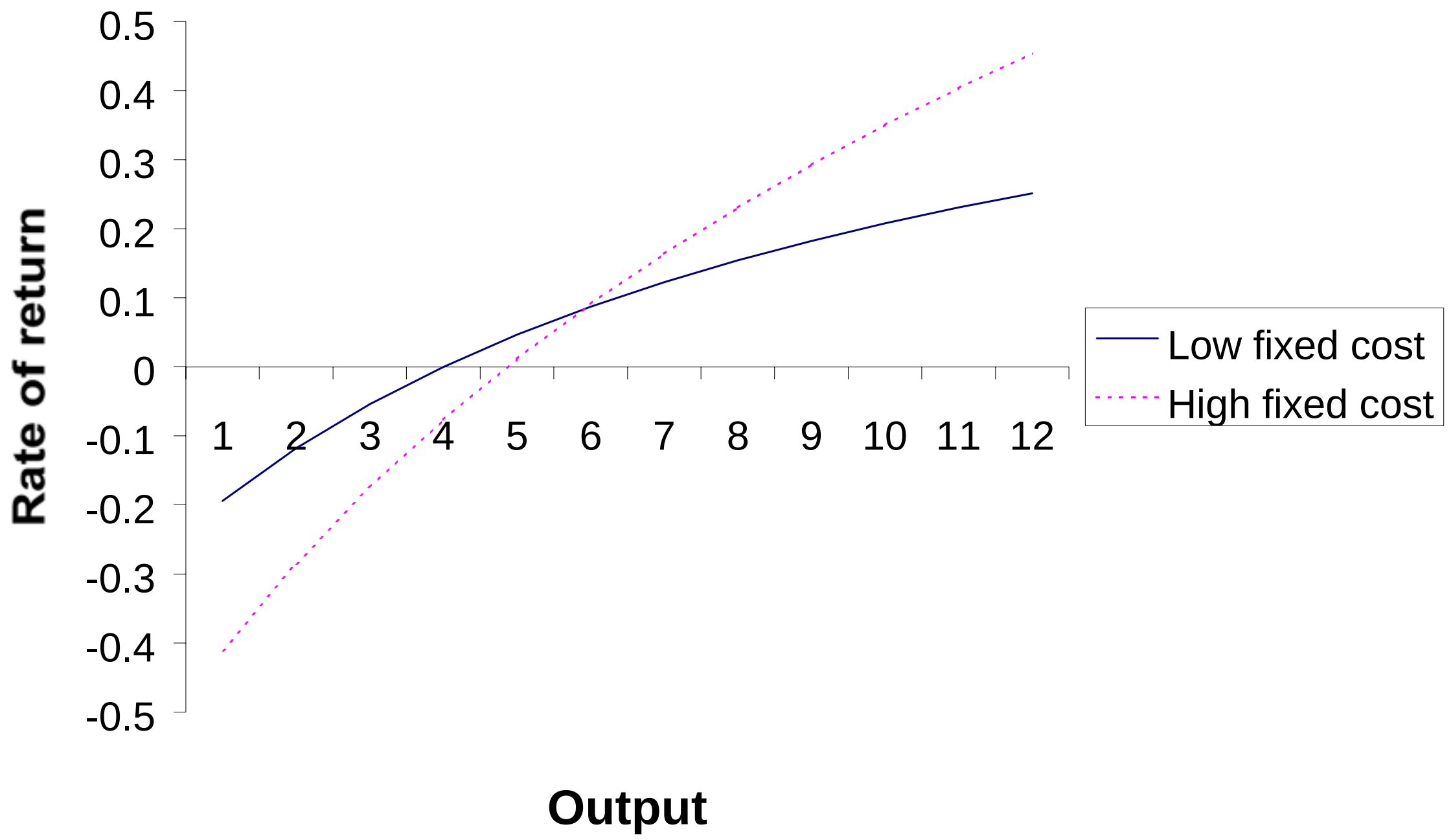
0.5
0.4
0.3
0.2
0.1
0
-0.1
-0.2
-0.3
-0.4
-0.5
Rate of return
1 2 3 4 5 6 7 8 9 10 11 12
Low fixed cost
High fixed cost
Output

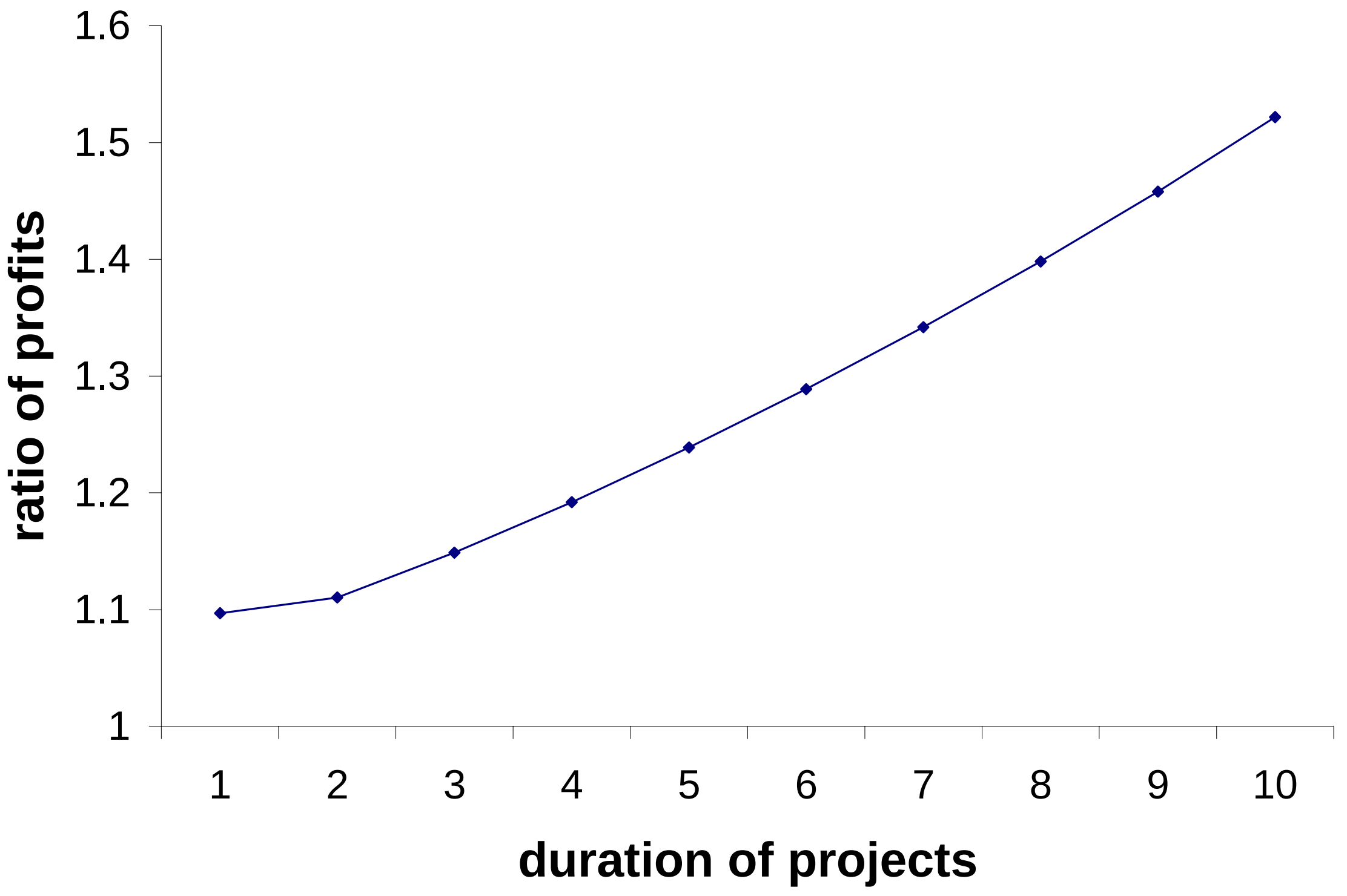

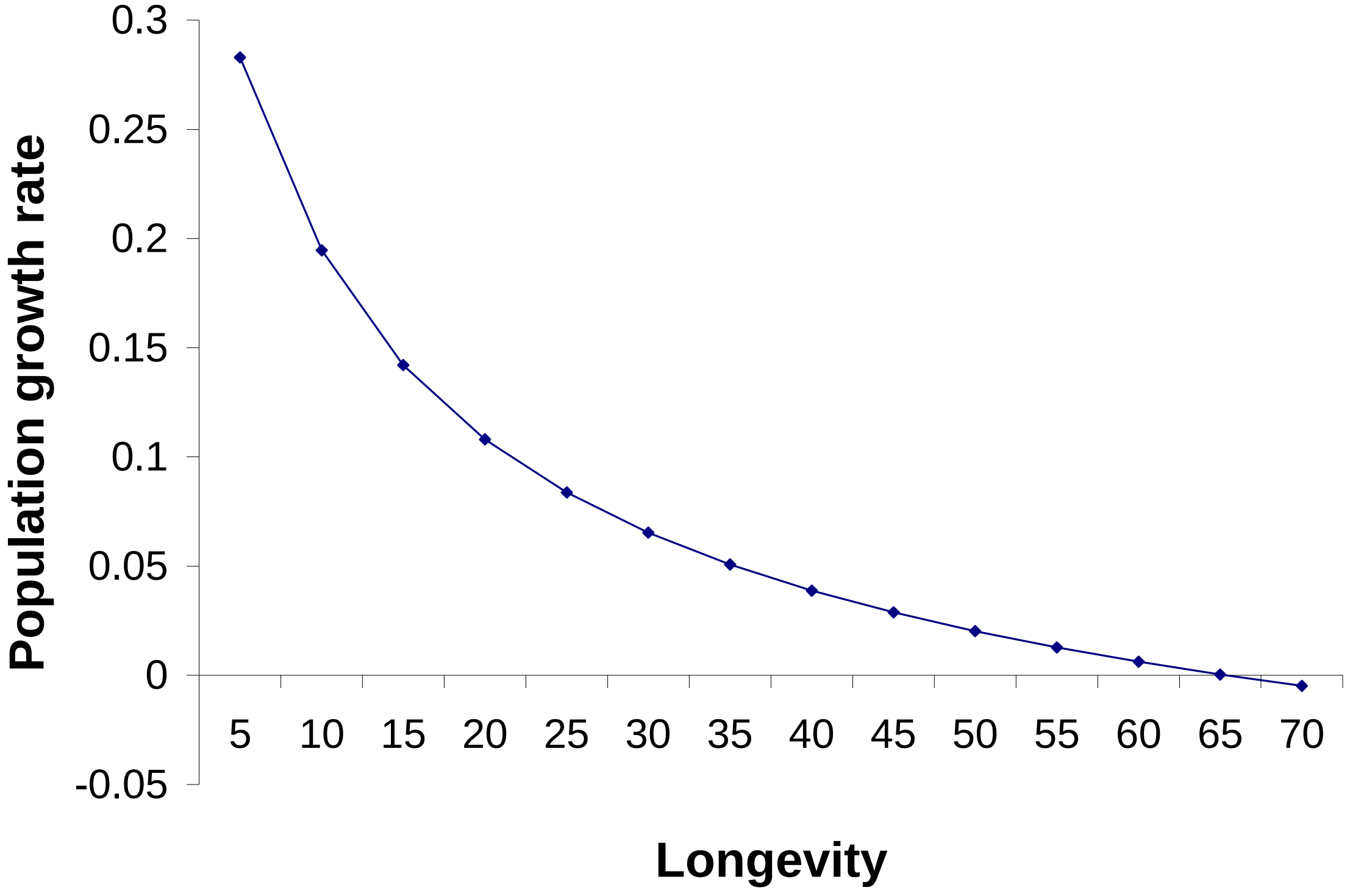
Population growth rate
0.3
0.25
0.2
0.15
0.1
0.05
0
-0.05
5 10 15 20 25 30 35 40 45 50 55 60 65 70
Longevity

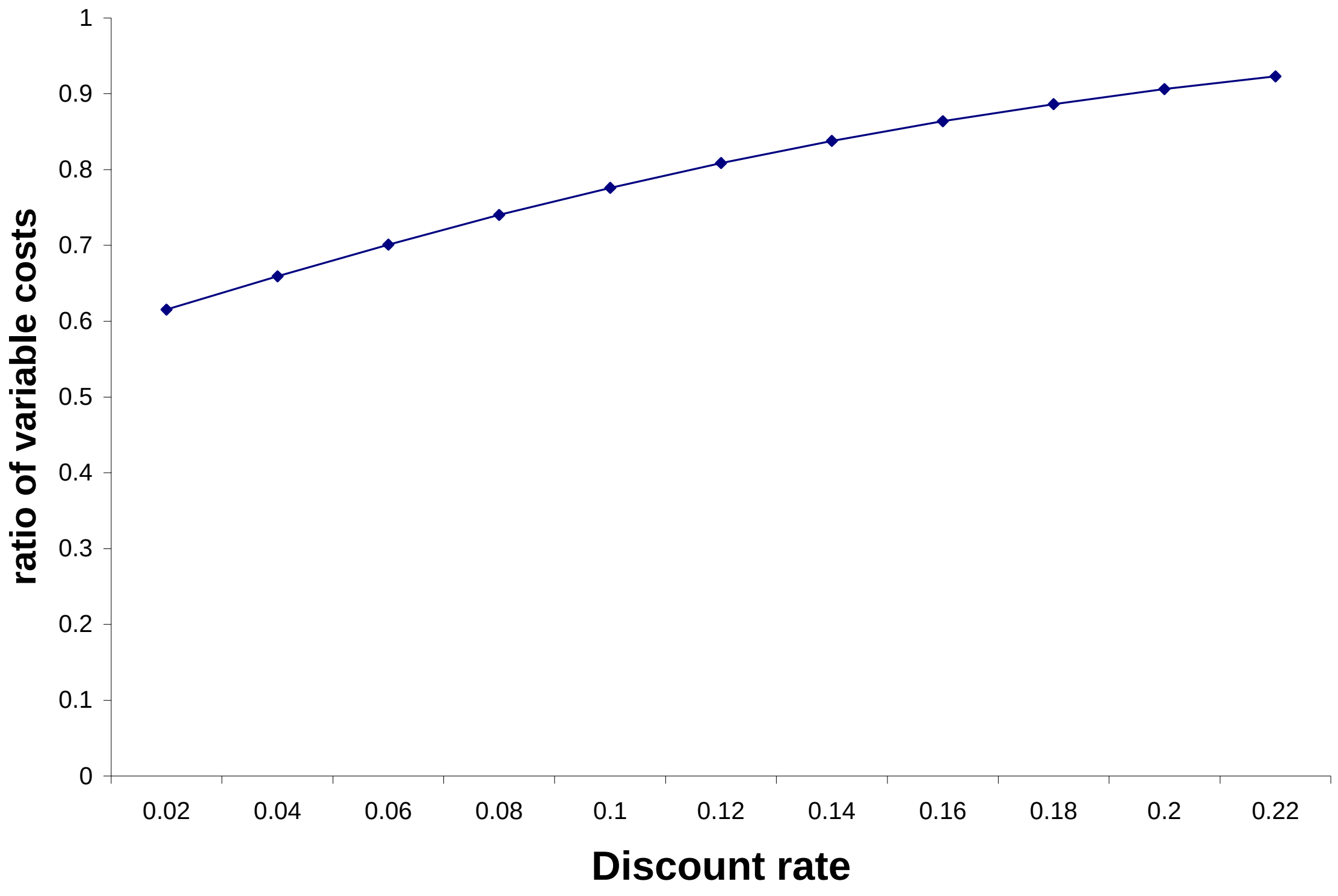

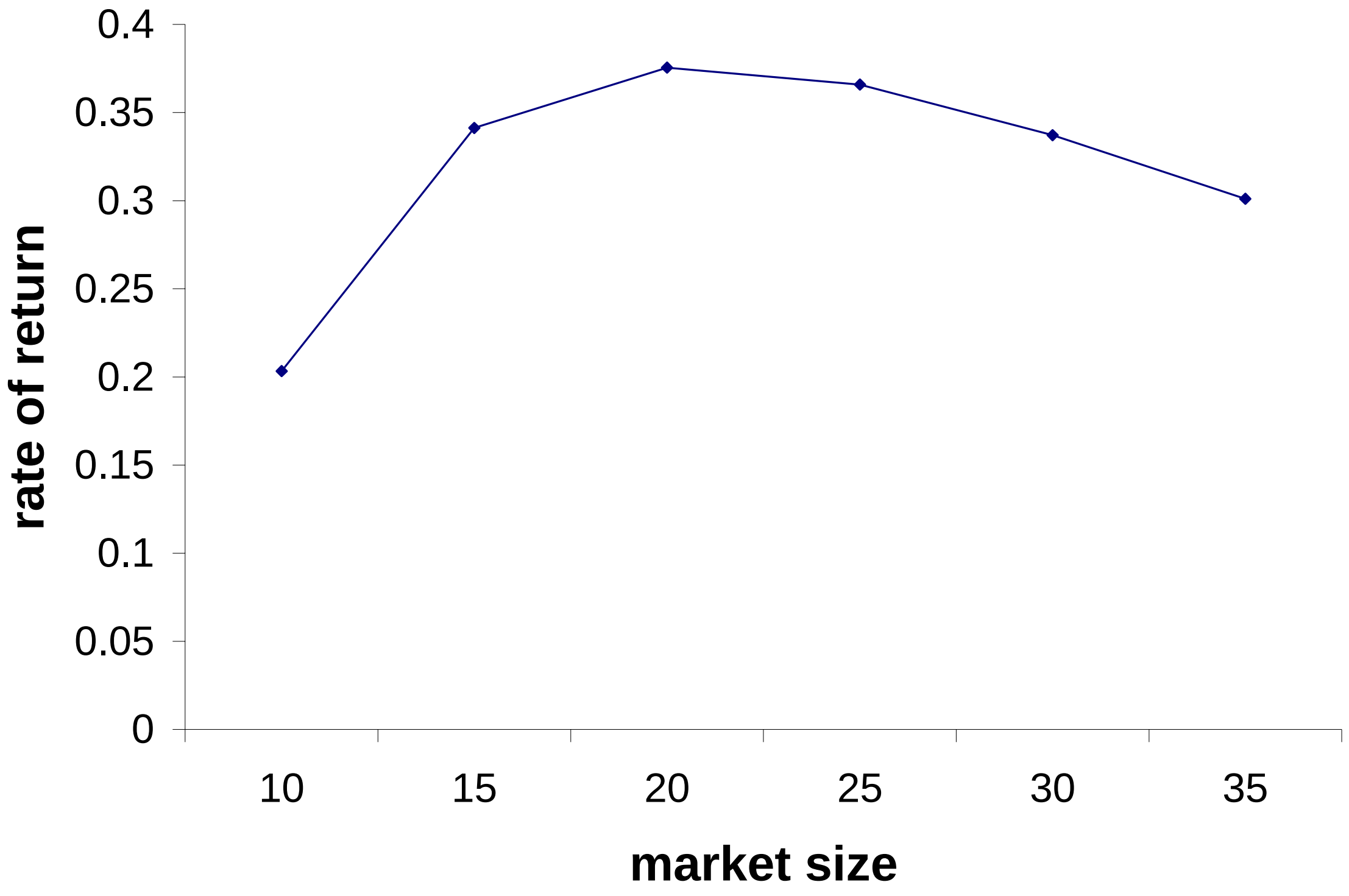